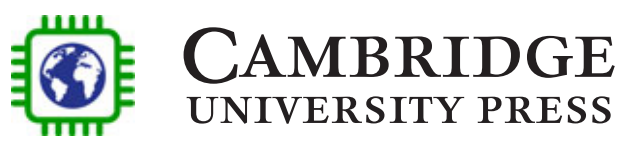

# Opportunities and challenges of quantum computing for climate modeling

Mierk Schwabe[1], Lorenzo Pastori[1], Inés de Vega[2], Pierre Gentine[3], Luigi Iapichino[4], Valtteri Lahtinen[5], Martin Leib[2], Jeanette Miriam Lorenz[6,7] and Veronika Eyring[1,8]

[1]Deutsches Zentrum für Luft- und Raumfahrt (DLR), Institut für Physik der Atmosphäre, Oberpfaffenhofen, Germany
[2]IQM Germany GmbH, München, Germany
[3]Center for Learning the Earth with Artificial Intelligence and Physics (LEAP), Columbia University, New York, NY, USA
[4]Quantum Computing and Technologies Department, Leibniz-Rechenzentrum der Bayerischen Akademie der Wissenschaften (LRZ), Garching b. München, Germany
[5]Quanscient Oy, Tampere, Finland
[6]Fraunhofer-Institut für Kognitive Systeme IKS, München, Germany
[7]Faculty of Physics, Ludwig-Maximilians-Universität München, München, Germany
[8]University of Bremen, Institute of Environmental Physics (IUP), Bremen, Germany
**Corresponding author:** Mierk Schwabe; Email: mierk.schwabe@dlr.de

## Abstract

Adaptation to climate change requires robust climate projections, yet the uncertainty in these projections performed by ensembles of Earth system models (ESMs) remains large. This is mainly due to uncertainties in the representation of subgrid-scale processes such as turbulence or convection that are partly alleviated at higher resolution. New developments in machine learning-based hybrid ESMs demonstrate great potential for systematically reduced errors compared to traditional ESMs. Building on the work of hybrid (physics + AI) ESMs, we here discuss the additional potential of further improving and accelerating climate models with quantum computing. We discuss how quantum computers could accelerate climate models by solving the underlying differential equations faster, how quantum machine learning could better represent subgrid-scale phenomena in ESMs even with currently available noisy intermediate-scale quantum devices, how quantum algorithms aimed at solving optimization problems could assist in tuning the many parameters in ESMs, a currently time-consuming and challenging process, and how quantum computers could aid in the analysis of climate models. We also discuss hurdles and obstacles facing current quantum computing paradigms. Strong interdisciplinary collaboration between climate scientists and quantum computing experts could help overcome these hurdles and harness the potential of quantum computing for this urgent topic.

### Impact Statement

Recently, quantum computing has been making rapid progress, with the first demonstrations of quantum advantage for selected problems. At the same time, climate change is becoming increasingly severe, and robust local information provided by climate models becomes more crucial. In this position paper, we explore how quantum computing methods could potentially help improve and accelerate climate models, and which obstacles remain.

## 1. Introduction

According to the Intergovernmental Panel on Climate Change (IPCC) Sixth Assessment Report (Masson-Delmotte et al., 2021), the effects of human-induced climate change are already felt in every region across the globe (Eyring et al., 2021a). There is an urgent need for better climate models that make regional projections possible and thus allow for more precise efforts at mitigation and adaptation (Shokri et al., 2022). Climate models do improve with each generation (Bock et al., 2020), however, systematic biases compared with observations still remain due to the limited horizontal resolution of the models, typically of the order of tens of kilometres (Eyring et al., 2021b). Models with a horizontal resolution of a few kilometres can explicitly represent deep convection and other dynamical effects (Hohenegger et al., 2020) and thus alleviate a number of biases (Sherwood et al., 2014), but have high computational costs. Even considering the expected increase in computing power (Ferreira da Silva et al., 2024; Stevens et al., 2024), a hierarchy of ideally hybrid ESMs, incorporating machine learning (ML) methods and physical modelling, will continue to be required (Eyring et al., 2024b). It is thus imperative to take advantage of novel technologies to both improve and accelerate climate models.

Quantum computers provide alternative computing paradigms, and have seen tremendous progress in the last years, see Figure 1. Size and quality of quantum hardware are steadily increasing, as well as the number of proposed quantum algorithms (Sevilla and Riedel, 2020), and a few experiments have claimed to have achieved quantum advantage (Arute et al., 2019; Lau et al., 2022; Madsen et al., 2022; Zhu et al., 2022; King et al., 2024). On the algorithmic side, a growing number of methods targeted to current devices are being developed and implemented, and new applications are envisioned.

In this perspective, we discuss how we foresee leveraging the potential of quantum computing in the context of climate modeling. First, we give an introduction to data-driven, ML-based climate modeling and to quantum computing. Then, we discuss the potential of quantum computers for climate modeling, especially pointing out algorithms available for current noisy intermediate-scale quantum (NISQ) devices. Finally, we discuss the next steps towards developing a climate model improved with quantum computing.

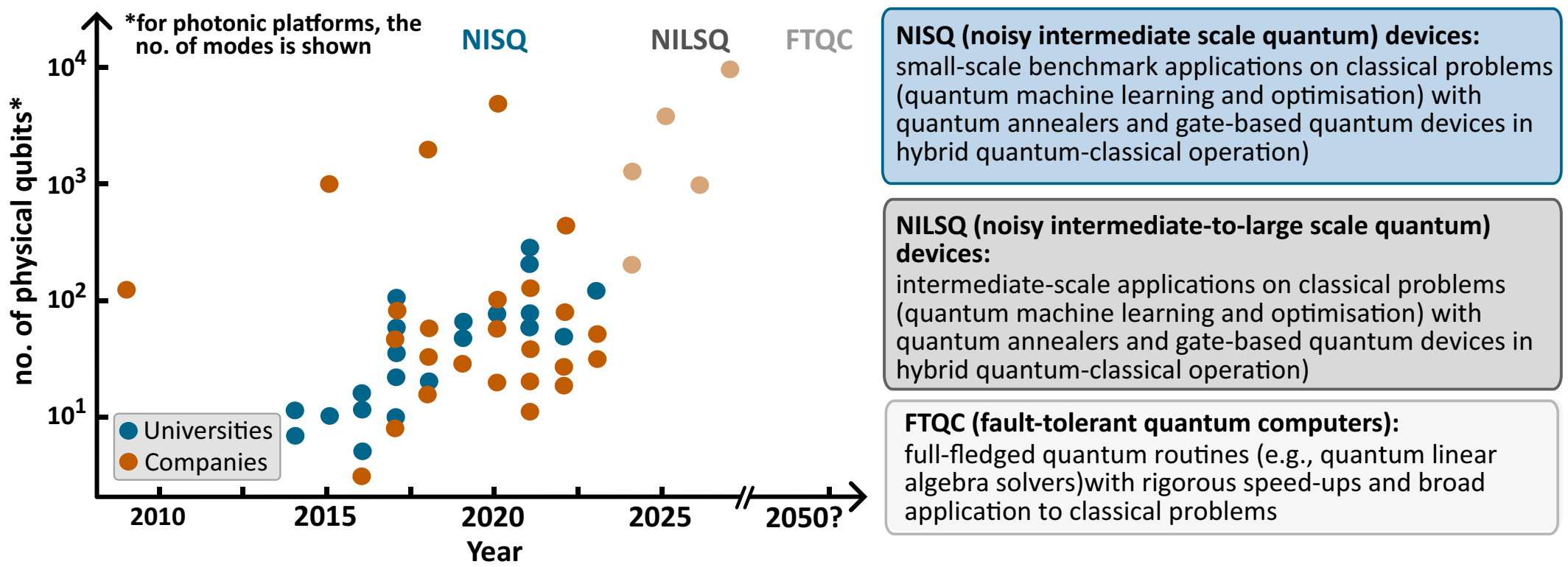

***Figure 1.*** *Evolution of the number of physical qubits of quantum computing and quantum simulation platforms by several companies and university research groups. The underlying data was collected from online sources and journal publications(D:Wave, 2023; IBM, 2023; PASQAL, 2023; rigetti, 2023; Google QAI, 2023; AQT, 2023; Quantinuum, 2023; Zhong et al., 2020; Scholl et al., 2021; Wu et al., 2021; Semeghini et al., 2021; Joshi et al., 2022; Ebadi et al., 2022). The shaded data points for future years were collected from the companies' roadmaps (D:Wave, 2023; PASQAL, 2023; IBM, 2023; IonQ, 2023).*

## 2. Climate modeling

Climate models are three-dimensional models based on fundamental laws of physics (Jacobson, 2005), see Box 1. The atmosphere is discretized over a horizontal grid covering the surface of the Earth, and vertical columns above each grid cell. In each grid box, state variables describe the physical properties,

***Box 1.*** *Fundamental equations of earth system models (ESMs)*

ESMs describe the evolution of the Earth system with time, given external forcings such as solar radiation and anthropogenic influences. They consist of coupled models of, e.g., the atmosphere, ocean, and land, Figure 2. For example, the state of the atmosphere is described by the equation of state

$$p = \rho R T \tag{1}$$

with pressure $p$, density $\rho$, the gas constant $R$, and temperature $T$. The barometric law describes the variation of pressure with altitude $z$

$$p(z) \approx p(0) \exp\left(-\frac{z}{H}\right) \tag{2}$$

with the atmospheric scale height $H$.

The dynamical properties of the atmosphere determine the evolution of mass, momentum, and energy. Mass conservation is expressed as the continuity equation

$$\frac{\partial \rho}{\partial t} + \nabla \cdot (\rho \boldsymbol{v}) = 0 \tag{3}$$

with the three-dimensional wind $\boldsymbol{v}$. The conservation of momentum in the system rotating with angular velocity $\Omega$ is described by the Navier–Stokes equation

$$\frac{\partial \boldsymbol{v}}{\partial t} + (\boldsymbol{v} \cdot \nabla)\boldsymbol{v} = -2[\Omega \times \boldsymbol{v}] - \frac{1}{\rho}\nabla p + \boldsymbol{g} + \nu \nabla^2 \boldsymbol{v} \tag{4}$$

with apparent gravitational acceleration $\boldsymbol{g}$ and kinematic viscosity $\nu$. Finally, the first law of thermodynamics results in the conservation of energy

$$c_V \frac{dT}{dt} + p \frac{d}{dt}\left(\frac{1}{\rho}\right) = Q \tag{5}$$

with the specific heat at constant volume $c_V$ and the diabatic heating term $Q$, which can be driven by absorption and emission of radiative energy or condensation and evaporation of water.

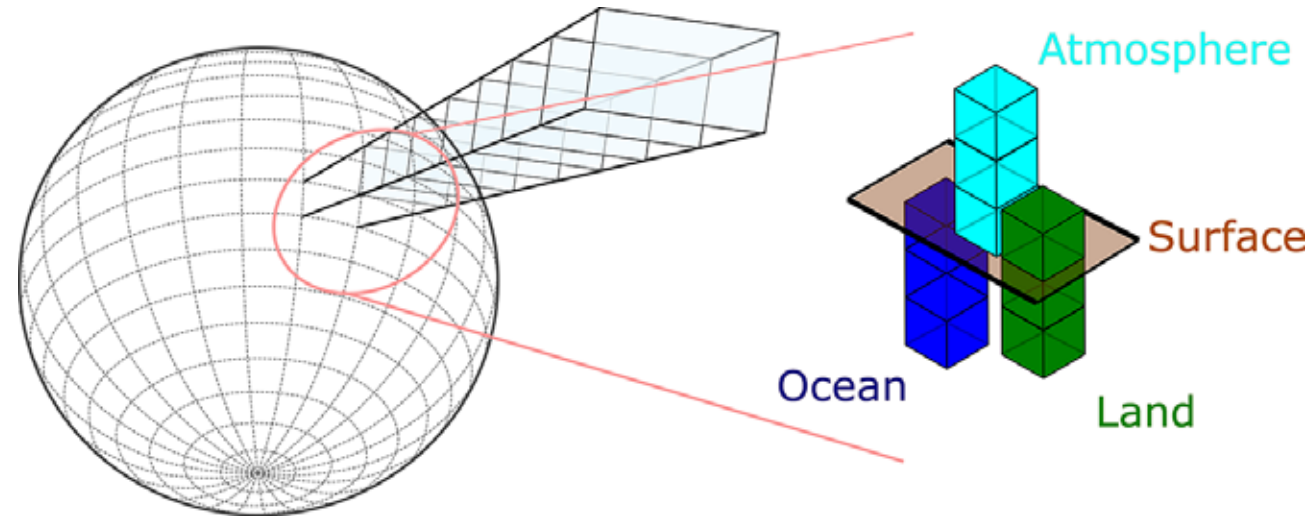

***Figure 2.*** *Schematic of an Earth system model (ESM). The ESM represents the state of the atmosphere, ocean and sea ice, and land using a grid covering the globe. For each component, physical properties such as water content in the atmosphere and soil or salinity of the ocean, the kinetic energy contained in the wind and currents, and the thermal energy contained in the temperature are represented. Following Gettelman and Rood (2016).*

see Box 1. During a time step of the simulation, the evolution of energy and mass and the motions of air and tracers are solved. Earth system models (ESMs) simulate the interactive carbon and other biogeochemical cycles in addition to the atmosphere, land, ocean, and sea ice physical states (Eyring et al., 2021b). Climate models and ESMs can simulate the mean state of the system, as well as natural variability, and how it may change given an external forcing (e.g., increasing the concentration of greenhouse gases). Not all relevant processes can be described by solving the fundamental equations, either because the resolution is not high enough (e.g., for shallow convection) or because the processes are not described by the equations (e.g., the formation of clouds and their influence on radiation transfer). Parameterizations represent the effects on the grid scale of the unresolved (subgrid-scale) processes as a function of the coarse-scale state variables. There have been many attempts to develop kilometre-resolution models that require fewer parameterizations and produce better input states for the remaining ones (Neumann et al., 2019; Stevens et al., 2019, 2024), yet even these do not eliminate the need for running ensembles of climate models nor can they completely resolve all key processes (e.g., shallow clouds). Besides the large computational costs, storing the output of high-resolution models is problematic. Even today, the cost and bandwidth of storage systems do not keep up with the available computing power (Schär et al., 2020). Especially the output of large ensembles of high-resolution climate models is impossible to store in its entirety, so that data needs to be coarse-grained or directly analysed while the simulations are running, and simulations need to be rerun when a specific analysis is required, trading storage for computation (Schär et al., 2020). Therefore, to enable high-resolution ensemble runs, there is an urgent need to accelerate climate models.

Even high-resolution simulations require the use of some parameterizations, such as for microphysics and turbulence, and these still cause biases (Stevens et al., 2019; Eyring et al., 2024b). These, as well as the parameterizations used in coarser-scale climate models, could be improved with ML methods (Bracco et al., 2025; Eyring et al., 2024b) that learn from short high-resolution model simulations or observations to represent processes that are unresolved by coarse climate models. Challenges remain (Eyring et al., 2024a) among which: 1) instabilities when ML-based parameterizations are coupled to the climate model, often due to the models learning spurious causal relationships (Brenowitz et al., 2020); 2) difficulty in generalizing beyond the training regime (Rasp et al., 2018), which is highly relevant in a changing climate, where the mean and extremes of climate variable distributions are shifting (Gentine et al., 2021). These challenges demonstrate the need for more expressive models (i.e., models that can learn a large variety of functions) that can be trained efficiently using potentially limited datasets.

All classical parameterizations as well as some data-driven ones (Grundner et al., 2024; Pahlavan et al., 2024), have parameters that need to be estimated to reduce the mismatch between observations and model results (Hourdin et al., 2017). This *tuning* is a very time-consuming process requiring considerable expert knowledge and computing time, which motivates the development of automatic algorithms to improve its efficiency and reproducibility (Hourdin et al., 2017; Bonnet et al., 2024). The tuned model should be evaluated against other climate models and against observations, in which novel techniques can be used to classify data sets and develop better-suited products (Kaps et al., 2023).

Summing up, there is an urgent need for faster and better climate models capable of running at high resolution, for more accurate and generalizable parameterizations, and for fast, reliable tuning and evaluation methods, also beyond classical ML algorithms.

## 3. Quantum computing: current status and challenges

The field of quantum computation deals with developing and controlling quantum systems to store and process information in ways that go beyond the capabilities of standard (classical) computers (Nielsen and Chuang, 2010), see Box 2. Quantum computers hold the promise of efficiently executing tasks intractable even for the largest supercomputer, including simulations of complex materials and chemicals, and solving optimization problems (Grumbling and Horowitz, 2019). Despite extraordinary theoretical and experimental developments in the last decades, we are still in the era of *NISQ devices*—— noisy intermediate-scale quantum devices counting up to few hundreds qubits, with several limitations caused

***Box 2.*** *Qubits, quantum gates, and entanglement*
The fundamental quantum information unit is the qubit. The state $|\psi\rangle$ of a qubit is a vector in a two-dimensional complex vector space, written as a superposition $|\psi\rangle = c_0|0\rangle + c_1|1\rangle$ of the two computational basis states $|0\rangle$ and $|1\rangle$ (Figure 3). Measuring the qubit in this basis yields 0 or 1 with probability $|c_0|^2$ or $|c_1|^2$, respectively, with subsequent collapse of $|\psi\rangle$ in the associated basis state. Qubits can be manipulated by means of quantum gates, which are unitary operators acting on the state vectors.

The superposition principle extends to the case of many qubits. The state of $N$ qubits has the form $|\psi\rangle = \sum_{\boldsymbol{\sigma}} c_{\boldsymbol{\sigma}} |\boldsymbol{\sigma}\rangle$, where $\boldsymbol{\sigma}$ denotes one of the $2^N$ possible bitstrings indexing the computational basis states, and $|c_{\boldsymbol{\sigma}}|^2$ the probability of obtaining $\boldsymbol{\sigma}$ as the outcome of a measurement in the same basis. An example of a two-qubit state has the form $|\psi\rangle = (|01\rangle + |10\rangle)/\sqrt{2}$. This is an example of an entangled state, where the qubits share non-local quantum correlations which result in correlated outcomes once they are measured (Nielsen and Chuang, 2010). Entanglement is generated using quantum gates that make selected qubits interact, and is a key ingredient for quantum computational advantage. The entanglement content of a quantum state is related to the computational complexity of representing it by classical means (Eisert et al., 2010), which generally requires computational resources scaling exponentially with the number of qubits $N$. Classically intractable quantum states can be prepared on quantum computers, and their properties can be measured. This translates to the ability of preparing and sampling from otherwise intractable probability distributions $|c_{\boldsymbol{\sigma}}|^2$, which can encode the solution to given problems.

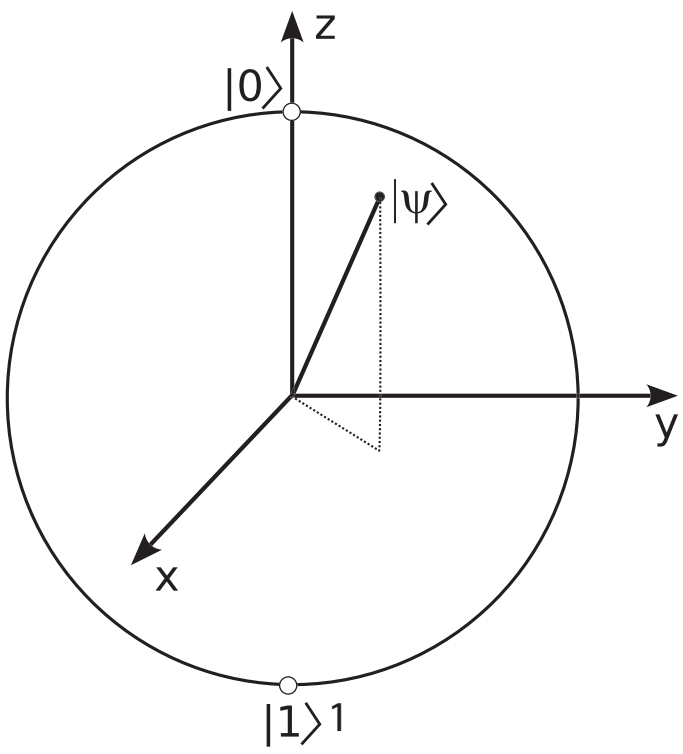

***Figure 3.*** *Representation of a qubit state as a vector on the Bloch sphere. The poles correspond to the basis states* $|0\rangle$ *and* $|1\rangle$.

by noise (Preskill, 2018). Despite these limitations, NISQ devices can already be used to tackle problems of academic interest, and it is foreseeable that quantum error correction methods will bring us fault-tolerant quantum computers in the future (Devitt et al., 2013; Campbell et al., 2017), see Figure 1. It is thus important to ask now whether the algorithms developed for quantum computers could help address the challenges faced by climate modelling (Singh et al., 2022; Nivelkar et al., 2023; Otgonbaatar et al., 2023; Rahman et al., 2024; Bazgir and Zhang, 2024). In the following, we review several relevant quantum computing paradigms and algorithms.

### 3.1. Quantum linear algebra solvers

Quantum linear algebra solvers make use of the fact that quantum computation using $N$ qubits is mathematically described by linear operators in vector spaces of (large) dimension $2^N$. Linear problems of size $M$ can be encoded in quantum states and operators using only logarithmically many qubits (in $M$). Provided this encoding, quantum solvers offer exponential improvements in terms of resources needed to solve a problem, requiring only $\mathcal{O}(\text{polylog}M)$ qubits and gates compared to the $\mathcal{O}(\text{poly}M)$ operations required for classical algorithms (Harrow et al., 2009). They can therefore significantly speed up large matrix operations, e.g., in the resolution of (partial) differential equations using finite difference/elements methods (Berry, 2014; Lloyd et al., 2020; Li et al., 2023). This exponential speedup depends on the ability of *efficiently* (i.e., with costs scaling polynomially in the number of qubits) encoding the problem data in the states and operators on the quantum device, and on the efficient readout of the properties of interest from the quantum state encoding the solution. These challenges, together with the limitations due to noise, restrict the current applicability of these routines to small-scale problems (Cai et al., 2013; Barz et al., 2014; Zheng et al., 2017).

### *3.2. Quantum annealing for optimization problems*

Quantum annealing offers a way of solving optimization problems on quantum devices (Das and Chakrabarti, 2008; Albash and Lidar, 2018). Quantum annealers address combinatorial problems with a discrete solution space. The solution to a given problem is encoded in the ground state of an Ising Hamiltonian (Das and Chakrabarti, 2008). This Hamiltonian is then realized on a quantum device, the annealer, and its ground state prepared by slowly steering an initial state towards it. In this way, the implemented quantum state globally explores the optimization landscape before tunnelling towards the optimal solution. Quantum annealers consisting of thousands of qubits are already available for use in academic and industrial applications (Yarkoni et al., 2022). This approach is potentially scalable to large problems, and the first claims of quantum advantage with quantum annealers have been made (King et al., 2024). However, noise is an open problem for quantum annealers (also due to the lack of fully fault-tolerant quantum annealing schemes (Pudenz et al., 2014)).

### *3.3. Parameterized quantum circuit models*

Parameterized quantum circuits (PQCs) are sequences of quantum gates depending on tunable parameters $\boldsymbol{\theta}$ that are optimized for the quantum device to solve a given problem. Applications include variational quantum eigensolvers (Peruzzo et al., 2014), quantum approximate optimization algorithms (Farhi et al., 2014), and quantum machine learning (QML) (Schuld and Petruccione, 2018; Schuld et al., 2015; Biamonte et al., 2017; Dunjko and Briegel, 2018; Schuld and Petruccione, 2018; Cerezo et al., 2022). PQCs are NISQ-friendly due to their limited depth, supplemented by the optimization of the parameters $\boldsymbol{\theta}$ that is achieved iteratively in a hybrid quantum-classical manner (Cerezo et al., 2021; Bharti et al., 2022): a cost function is measured on the quantum device and fed to a classical routine that proposes new $\boldsymbol{\theta}$ for the next iteration. This results in shorter and classically optimized circuits that can run within the coherence time of NISQ devices.

In the context of QML, PQCs find applications in regression, classification, and generative modeling tasks (Cerezo et al., 2022). In *regression* and *classification*, PQCs are used as function approximators and are often referred to as quantum neural networks (QNNs) (Farhi and Neven, 2018). Classical input data $\boldsymbol{x}$ first is encoded in a quantum state $|\phi(\boldsymbol{x})\rangle$. Then the output is calculated as the expectation value of an observable in the output state $|\psi(\boldsymbol{x};\boldsymbol{\theta})\rangle = \widehat{U}_{\boldsymbol{\theta}}|\phi(\boldsymbol{x})\rangle$, where $\widehat{U}_{\boldsymbol{\theta}}$ denotes the action of the PQC. The embedding $\boldsymbol{x} \rightarrow |\phi(\boldsymbol{x})\rangle$ needs to be carefully chosen as it strongly influences the model performance (Schuld and Petruccione, 2018; Pérez-Salinas et al., 2020; Schuld et al., 2021), and potentially requires compression strategies (Dilip et al., 2022). The higher expressivity of QNNs (Du et al., 2020, 2021; Yu et al., 2023b), makes them interesting ansatzes for ML tasks. Furthermore, several works have investigated their generalization properties, with promising results (Cong et al., 2019; Abbas et al., 2021; Caro et al., 2022, 2023). Regarding their trainability, while suitably designed QNNs show desirable geometrical properties that may lead to faster training (Abbas et al., 2021), in general, the training of PQCs can be hindered by the presence of barren plateaus (Ragone et al., 2024) and potentially requires advanced initialization strategies (Zhang et al., 2022). However, in quantum convolutional neural networks (QCNNs), barren plateaus do not seem to be a problem (Pesah et al., 2021).

The embedding $\boldsymbol{x} \to |\phi(\boldsymbol{x})\rangle$ can be thought of as a feature map from the input $\boldsymbol{x}$ to the (large) Hilbert space of quantum states (Schuld and Killoran, 2019; Havlíček et al., 2019). This observation constitutes the basis of *quantum kernel methods* (Schuld and Killoran, 2019; Havlíček et al., 2019), where quantum kernels are constructed from inner products $\langle \phi(\boldsymbol{x}) | \phi(\boldsymbol{x}') \rangle$ measured on the quantum device. The constructed kernels are then used in subsequent tasks, e.g., in data classification (Havlíček et al., 2019), or Gaussian process regression (Otten et al., 2020; Rapp and Roth, 2024). Despite their advantage on specific datasets (Huang et al., 2021), quantum kernel methods may also suffer from trainability issues analogous to barren plateaus in QNNs (Thanasilp et al., 2022), which highlights the requirement of a careful design.

PQCs also find a natural application as *generative models* (Amin et al., 2018; Dallaire-Demers and Killoran, 2018; Coyle et al., 2020). The output state of a PQC indeed corresponds to a probability distribution over an exponentially large computational basis, via the relation $P_{\boldsymbol{\theta}}(\boldsymbol{\sigma}) = |\langle \boldsymbol{\sigma} | \widehat{U}_{\boldsymbol{\theta}} | \phi_0 \rangle|^2$, with $\boldsymbol{\sigma}$ indexing the computational basis states, and $|\phi_0\rangle$ a predefined reference state. For generative tasks, the parameters $\boldsymbol{\theta}$ are optimized for $P_{\boldsymbol{\theta}}$ to approximate the probability distribution underlying a set of training data. The PQC is then used to sample the distribution $P_{\boldsymbol{\theta}}$. These models have improved representational power over standard generative models (Du et al., 2020), but also require careful design to avoid trainability issues (Rudolph et al., 2023).

We finally remark that despite the several hints to potential advantages on specific tasks, it is not yet clear whether a practical advantage of QML on real-world classical datasets can be demonstrated. Among the practical hurdles are the aforementioned trainability issues, as well as the necessity of efficient coupling with HPC and the inevitable presence of noise.

### 3.4. Integrated high-performance quantum computing

PQCs are based on the interplay between a quantum and a classical component within a hybrid algorithm. NISQ systems thus need to be integrated with classical computing resources at the hardware level. As quantum circuits become more complex, the classical component in PQCs and thus the required computing resources will grow as well, to the scale of an HPC problem. Given the trend towards heterogeneous development of modern HPC systems, e.g., increasingly using hardware accelerators like graphics processing units (Schulz et al., 2021), this leads to consider quantum hardware as a new class of accelerators for dedicated tasks within HPC workflows (Humble et al., 2021; Rüfenacht et al., 2022). The high performance computing—quantum computing (HPCQC) integration is an interdisciplinary challenge with aspects at the hardware, software, programming, and algorithmic level. The hardware integration is crucial to reduce latencies, especially in iterative variational algorithms (Rüfenacht et al., 2022). Moreover, the development of a single software stack for integrated systems, including the offloading of tasks to quantum accelerators and the scheduling of those resources, is necessary for ensuring operation and a smooth user experience (Schulz et al., 2022).

The HPCQC integration is not a concept limited to the NISQ development phase of quantum hardware, but is also relevant for future, fault-tolerant systems. At that point, we expect that the HPC computational resources will have to take over further tasks related to the operation of the quantum hardware, like circuit compilation and error control (Davenport et al., 2023; Maronese et al., 2022).

### 3.5. Error correction and mitigation

Current NISQ devices are still modest in qubit counts (tens to hundreds), circuit depths (accommodating up to a thousand gate operations), and coherence times (from microseconds to seconds, depending on the platform (Byrd and Ding, 2023)). The achievable circuit depth and the repetition rate influence the complexity and the practical applicability of QML models, respectively. Hardware noise may negatively impact QML trainability and is also one of the main limiting factors for quantum annealing and for fault-tolerant applications.

Achieving fault-tolerance is the goal of *quantum error correction* methods, in which one logical qubit is represented using several physical ones, to detect and correct errors at runtime (Devitt et al., 2013). This is a crucial requirement for quantum algorithms relying on extensive numbers of quantum operations, such as quantum linear algebra subroutines. Given the substantial resource overhead needed for quantum error correction (Davenport et al., 2023), its practical implementation can be considered in its infancy despite recent rapid advances (Google Quantum AI and Collaborators, 2025).

Other techniques to reduce noise-induced biases, without correcting for faulty qubits or operations, go under the name of *error mitigation* and are based on post-processing measurements collected from a suitably defined ensemble of quantum computation runs (Cai et al., 2023). Many mitigation techniques are currently being researched, such as zero-noise extrapolation and probabilistic error cancellation (Cai et al., 2023). These are particularly relevant for NISQ-targeted applications, as they can be implemented with little overhead in the number of gates and qubits. Hence, they are promising for “de-noising” the predictions from PQC-based methods.

## 4. Quantum computing for climate modeling

We now move on to considering the potential that the quantum algorithms presented in the previous sections can have for climate modelling. We discuss potential applications in solving the underlying differential equations, in model tuning, analysis, and evaluation, and in improving subgrid-scale parameterizations (Figure 4).

### *4.1. Accelerating the resolution of high-dimensional differential equations*

Solving differential equations is at the root of climate modelling, especially in the atmosphere and ocean components. Dynamical conservation laws are represented by PDEs, and chemical reactions are described by ordinary differential equations (ODEs) (Alvanos and Christoudias, 2019; Sander et al., 2005; Zlatev et al., 2022). The resources needed for solving them grow rapidly with increasing resolution of the models. This calls for methods for efficiently encoding and processing the climate state variables in all the model cells. Several quantum algorithms seeking an exponential reduction of the needed resources have been recently proposed in the context of computational fluid dynamics (Gaitan, 2020; Steijl and Barakos, 2018).

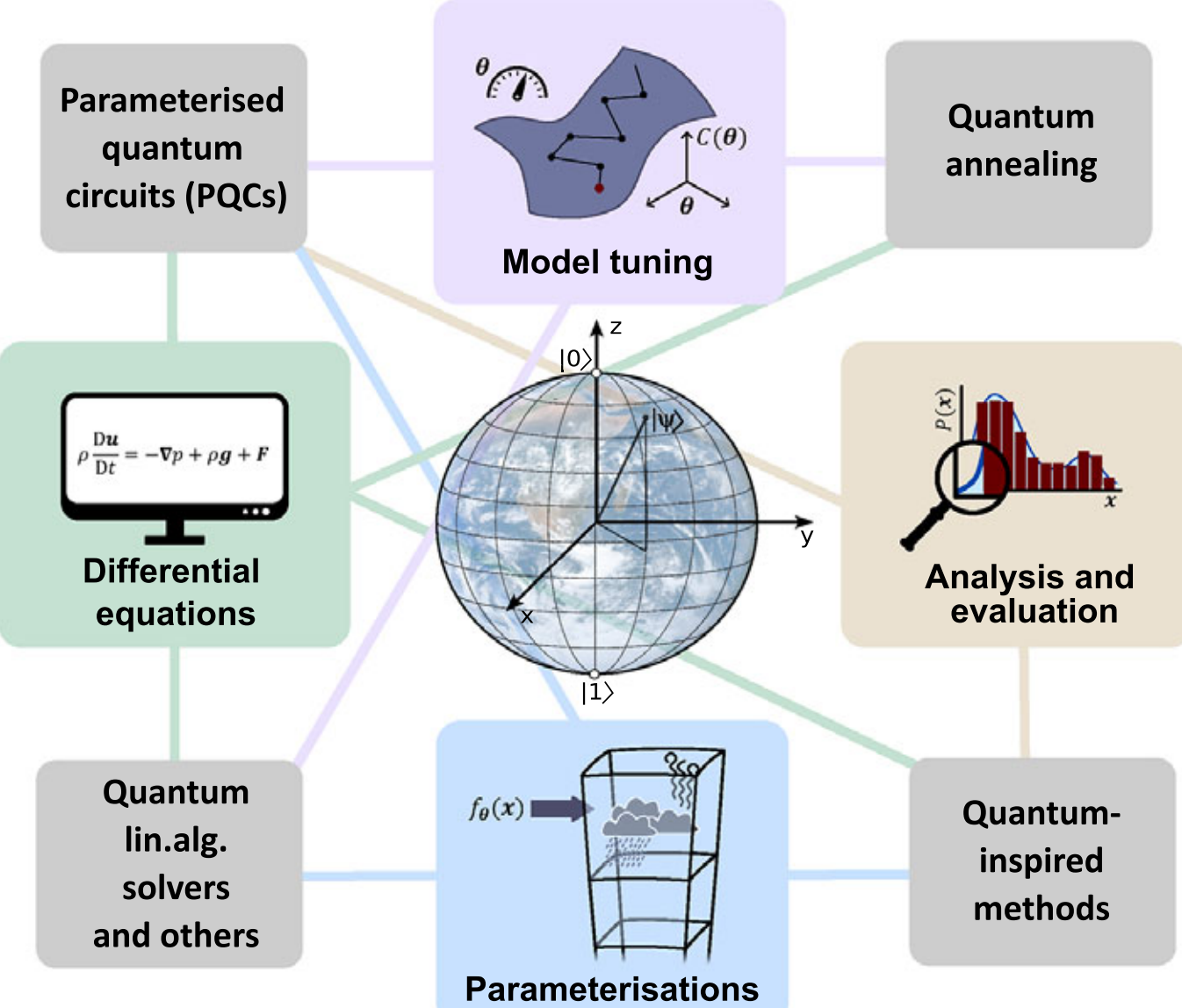

***Figure 4.*** *Overview of the climate modeling tasks and the category of matching quantum computing algorithms. (Image of the Earth by NASA/Apollo 17).*

Methods for time-dependent fluid-flow problems have been developed in the framework of gate-based quantum computing (Steijl and Barakos, 2018; Mezzacapo et al., 2015; Gaitan, 2020) and quantum annealers (Ray et al., 2019). For example, using spatial discretization of the PDEs, Gaitan (2020) reduces the one-dimensional Navier–Stokes equations to a system of ODEs

$$\frac{d\boldsymbol{U}}{dt} = f(\boldsymbol{U}), \tag{6}$$

with $\boldsymbol{U}$ being a vector depending on $\rho$, $v$, and $T$, and the driver function $f(\boldsymbol{U})$ depending on the discretization procedure. Kacewicz (2006) showed that an almost optimal quantum algorithm exists for finding an approximate solution to a set of nonlinear ODEs. As shown by Gaitan (2020), the solution of Eq. (6) is approximated by discretizing it in both the spatial and time domains, and then determining the approximate solution using quantum amplitude estimation (QAEA) (Brassard et al., 2002). These hybrid approaches for Navier–Stokes equations (Gaitan, 2020) could thus potentially achieve up to exponential speed-up compared to deterministic classical algorithms, due to the quantum subroutine (Brassard et al., 2002). However, the stability of such algorithms with respect to the spatial discretization still remains an open issue, and these methods inherit the challenges that quantum linear solvers may face, related to the data encoding and readout, and to the implementation of the linear operator defining the problem (Aaronson, 2015).

Quantum lattice–Boltzmann methods (QLBM) offer alternatives to quantum linear solvers, circumventing solving linear systems directly (Budinski, 2021, 2022; Schalkers and Möller, 2022). The lattice–Boltzmann method, a mesoscopic stream-and-collide method for probability densities of fluid particles, lends itself to quantum solution natively and efficiently (Budinski et al., 2023; Li et al., 2023). QLBM algorithms can be fully quantum (Budinski, 2021), or hybrid, such as Budinski's Navier–Stokes algorithm (Budinski, 2022), where quantum-classical communication is needed to incorporate non-linearities. Other approaches for non-linearities in QLBM exist, e.g., by using Carleman linearization (Itani and Succi, 2022). Koopman operators can also be used to induce linearity (Bondar et al., 2019). Also, QLBM faces the challenge related to data readout and encoding: for some QLBM models, this must be repeated at all time steps, which makes efficient time-marching a problem of prime importance for QLBM.

Data-driven quantum-assisted approaches are also a possibility to predict the evolution of climate systems, as demonstrated by Jaderberg et al. (2024) in the context of weather models.

In summary, the encoding and readout of classical variables is one of the primary limitations that need to be addressed for applying the aforementioned routines to problems on scales of a climate model. Even though 30 logical qubits could be sufficient to store one billion climate variables (Tennie and Palmer, 2023) and to implement their time evolution using quantum circuits of polynomial depth, it is still unclear what type of climate states can be efficiently encoded in quantum states, and which properties of the climate can be efficiently measured. Although full quantum state reconstruction is not experimentally feasible for large quantum systems (it requires a number of measurements growing exponentially in the number of qubits (Struchalin et al., 2021)), fundamental insights on the properties that can be efficiently read out from quantum states exist in the framework of shadow tomography (Huang et al., 2020): these results will need to be adapted to the specific use-case of climate simulation and subsequent evaluation.

### *4.2. QML-based parameterizations*

QML models, such as QNNs, can be used to develop data-driven parameterizations by training them with data from short high-resolution simulations, as shown in Figure 5, in analogy to what is currently done with classical machine learning (Gentine et al., 2021; Eyring et al., 2024a). The properties of QML models (Yu et al., 2023b; Abbas et al., 2021; Caro et al., 2022; Huang et al., 2021) may lead to highly expressive parameterization models requiring less parameters, hence potentially requiring less training data if matched by good generalization capabilities. These attributes would be crucial for developing stable and reliable long-term climate projections. First quantum-enhanced ML algorithms for the emulation of climate model data have recently been demonstrated by Bazgir and Zhang (2024) using the ClimSim dataset (Yu et al., 2023a): Using a subset of the dataset, they used three different quantum models to predict target values such as snow and rain rates and various solar fluxes using surface pressure, insolation, and the surface latent and sensible heat fluxes as input. The quantum models they tested were a QCNN, a quantum multilayer perceptron (QMP), and a quantum encoder–decoder (QED), each compared to their classical counterparts. The quantum models typically outperformed the classical models. In another work (Pastori et al., 2025), some of the authors of the present manuscript recently demonstrated a QML-based parameterization of cloud cover for a climate model. The practical assessment of QML advantages in such tasks requires, however, addressing several open points. Encoding climate variables in quantum states constitutes a critical step, given the efficiency requirements in the NISQ era and the absence of quantum random access memory. Specifically tailored classical data-compression routines before the encoding stage in a QNN could include techniques based on feature selection (Mücke et al., 2023), variational encoding sequences (Behrens et al., 2022), or tensor networks, which can be naturally translated to quantum circuits (Dilip et al., 2022; Cichocki et al., 2016).

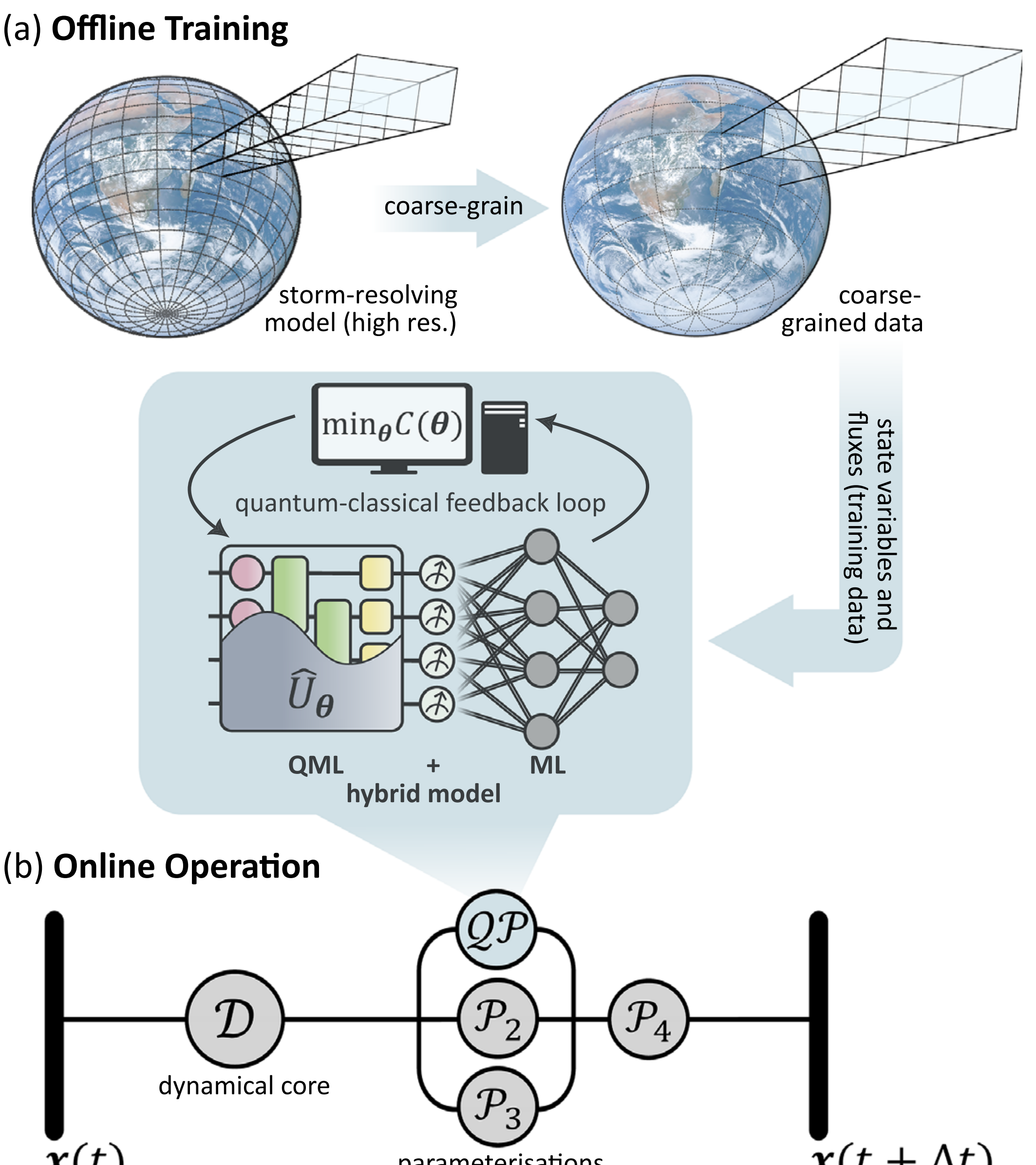

***Figure 5.*** *Hybrid quantum-classical approach for QML-based parameterizations. (a) Offline training: variables from cloud-resolving models are coarse-grained to the scale of the target climate model. The subgrid part of the variables of interest (e.g., fluxes) is calculated and used as a target for training QML-based parameterizations, possibly complemented with ML-based pre- and post-processing steps. The training of the QML model is typically assisted by a classical computer. The result is a replacement for a conventional parameterization and is coupled to the target climate model. (b) Processes occurring while the coarse-scale climate model (summarized by variables* $\boldsymbol{x}$*) is advanced from time step* $t$ *to* $t + \Delta t$*. First, the dynamical core* $\mathcal{D}$ *is run, followed in parallel or sequentially by the various parameterizations, denoted here with* $\mathcal{P}_2$ *to* $\mathcal{P}_4$*. The QML-parameterization* $\mathcal{QP}$ *is run online, replacing the corresponding traditional parameterization* $\mathcal{P}_1$*.*

The choice of the QNN structure is problem dependent and needs to match the hardware constraints. Also, it is desirable to encode conservation laws and physical constraints at the level of the underlying PQC, e.g., via the use of equivariant gate sets (Meyer et al., 2023), or by directly encoding the physical laws into the model (Markidis, 2022).

Another point to consider is the probabilistic nature of the measurement outcomes. While the inherent quantum noise could prove to have some benefits for climate simulations (Tennie and Palmer, 2023), the accuracy of the model prediction depends on the number of circuit evaluations, thus influencing the runtime. For nowadays' implementations, the runtime of such QNN parameterizations constitutes a limiting factor in their usages when coupled with the climate model, since they need to be run at each time-step for all model cells. Considering a time requirement of 10 μs per QNN run, and approximately 100 required measurement runs (Pastori et al., 2025), the evaluation of a QNN parameterization for a climate model with 1 million grid cells would take 1000 s per model time-step, which is clearly prohibitive. Expected advances in quantum hardware and its coupling with HPC will also increase the potential for integrating QML routines in climate models.

### 4.3. Improving climate model tuning

QML models and quantum optimization routines could also be used to improve the accuracy and speed of climate model tuning following an automatic tuning procedure (Figure 6). First, the tuning goals and parameters are chosen. Then, emulators are constructed to approximate the model output and to speed up the calibration process (Bellprat et al., 2012; Watson-Parris et al., 2021). Potentially optimal parameter regimes are inferred (Couvreux et al., 2021; Watson-Parris et al., 2021; Allen et al., 2022; Zhang et al., 2015; Cinquegrana et al., 2023; Bonnet et al., 2024), and new climate model outputs are evaluated in the proposed parameter regime. The procedure is iterated until the tuning goals are achieved.

The main quantum computing applications that we envision here concern the emulator construction and its optimization. Typical choices of *emulators in automatic tuning schemes* are neural networks and Gaussian processes (Watson-Parris et al., 2021). The former choice suggests the use of QNN emulators, while the latter option implies the choice of an underlying kernel function, suggesting the use of quantum kernels in Gaussian process regression (Otten et al., 2020). For example, Rapp and Roth (2024) develop a quantum Bayesian optimization algorithm based on quantum Gaussian processes. They construct a kernel by embedding data into the Hilbert space of a quantum system

$$|\phi(\boldsymbol{x};\boldsymbol{\theta})\rangle = U(\boldsymbol{x};\boldsymbol{\theta})|0\rangle. \tag{7}$$

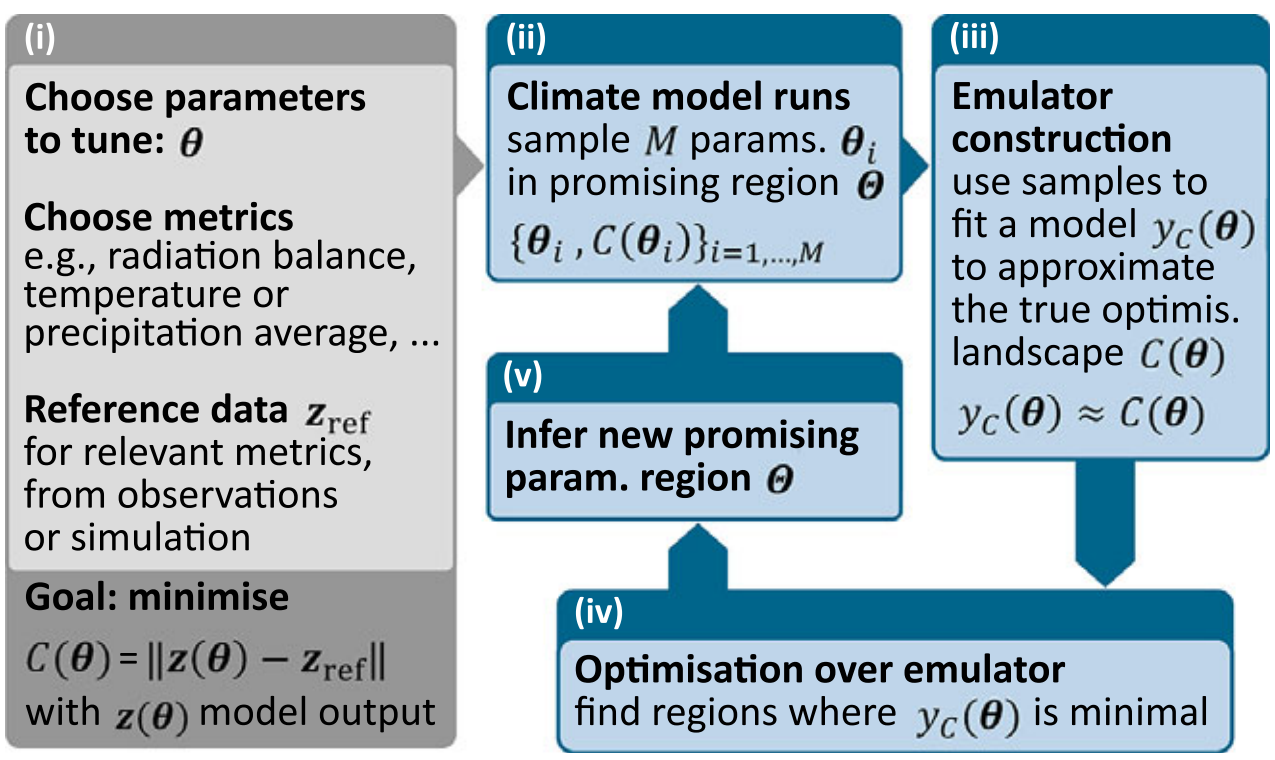

***Figure 6.*** *General steps of an automatic tuning protocol for climate models. In Step (i), the tuning goals and parameters are chosen. Then, in Step (ii) and (iii), emulators are constructed to approximate the model output and to speed up the calibration process. Potentially optimal parameter regimes are inferred in steps (iv) and (v), and new climate model outputs are evaluated in the proposed parameter regime. The procedure is iterated until the tuning goals are achieved.*

The unitary operator $U(\boldsymbol{x};\boldsymbol{\theta})$ encodes the classical data $\boldsymbol{x}$ into a quantum state and can depend on trainable parameters $\boldsymbol{\theta}$. It implements the quantum feature map $\phi$ and defines the density matrix $\rho(\boldsymbol{x}) = U(\boldsymbol{x})|0\rangle\langle 0|U^{\dagger}(\boldsymbol{x})$. The quantum kernel then, is given by

$$k(\boldsymbol{x},\boldsymbol{x}') = \mathrm{Tr}[\rho(\boldsymbol{x})\rho(\boldsymbol{x}')]. \tag{8}$$

The corresponding Gram matrix is then substituted as a covariance matrix into a classical Gaussian process, and the feature map is optimized using maximum likelihood estimation. The resulting algorithm is used for a hyperparameter optimization of a machine learning model.

Exploring the potential of QML for constructing emulators is of great interest since highly expressive models, if matched by good generalization properties, could benefit from fewer trainable parameters and training data, and hence fewer climate model simulations. The *optimization of the acquisition function* resulting from the emulator yields potentially optimal parameter sets and can be tackled with quantum heuristics such as quantum annealing, quantum-approximate optimization algorithms, or variational quantum eigensolvers, upon suitable discretization of the function to be optimized. Several works have recently explored the performance of these methods on continuous variable optimization problems (Izawa et al., 2022; Koh and Nishimori, 2022; Abel et al., 2022). Furthermore, continuous variable quantum computers, e.g., photonic platforms, could offer alternatives to avoid parameter space discretization (Enomoto et al., 2023).

### 4.4. *Assisting model analysis and evaluation*

Improved climate model analysis and evaluation routines lead to better identification of climate model biases. An important task towards these goals involves learning the probability distributions underlying the models' outputs. Classical generative methods are able to produce high-fidelity, realistic examples of climate data, e.g., with dynamics consistent with diurnal cycles (Besombes et al., 2021; Behrens et al., 2022). Given the ability of quantum systems to efficiently, i.e., with few parameters, describe complex probability distributions (Du et al., 2020; Wu et al., 2021; Hangleiter and Eisert, 2023), quantum generative models are natural candidates for achieving good extrapolating abilities using a small amount of training data. Quantum generative models can be built using PQCs, and trained either by minimizing the divergence between the target and the PQC distribution (Coyle et al., 2020) or in an adversarial manner (Dallaire-Demers and Killoran, 2018). Once trained, new samples following climate variable distributions can be efficiently generated and used for further analysis.

QML could also offer improvements in the classification of climate data, which is useful for improving and complementing observational products to subsequently use them for model evaluation (Kaps et al., 2023). PQCs again offer alternative ansatzes to classical ML with potential benefits (Du et al., 2020, 2021; Abbas et al., 2021; Caro et al., 2022). Existing QML methods for classification include quantum kernels and variants thereof (Schuld and Killoran, 2019; Havlíček et al., 2019; Huang et al., 2021), and QNNs (Farhi and Neven, 2018; Abbas et al., 2021; Hur et al., 2022). Among these, quantum convolutional neural networks (Cong et al., 2019) have shown remarkable trainability and generalization capabilities when used on quantum data (Cong et al., 2019; Pesah et al., 2021; Du et al., 2021), although recent works have suggested the possibility of their classical simulability (Cerezo et al., 2023).

## 5. Ways ahead

Building on ML-hybrid modeling, in this perspective, we highlight the potential of quantum computing and QML to address challenges in climate modeling. The potential benefits of these applications currently come with limitations that need to be overcome, particularly concerning the short coherence times in NISQ devices, the need for an efficient coupling to HPC facilities, the large amount of data needed for typical climate applications, and the limited capacity for read-out. Nevertheless, it is crucial to start exploring the applications proposed here, to timely adapt them to future quantum devices, and to enable co-design approaches bringing together hardware engineers, software developers, and potential users in the climate modelling community. In the following, we outline some possible first steps. For each task, the role of the climate modeling community is to provide simplified problem instances and models on which routines and solvers can be tested. From the quantum side, it is crucial to identify which problems most naturally lend themselves to quantum solutions and to critically assess the potential advantages of suitable quantum algorithms. Furthermore, a task for both communities, involving also the computer science community, is to efficiently couple quantum computers and HPC facilities.

Quantum algorithms for solving partial differential equations could be applied to *speed up the dynamical core of climate models.* However, due to the limited read-out capacities, only few variables could be extracted from these runs. Already, with current HPC systems, it is cheaper to rerun the model if additional analyses are needed instead of storing the entire output. Quantum computers might follow this trend to an extreme degree, resulting in only a very limited output for each run, but repeating the runs a large number of times. In the near future, work could be started by solving simple equations such as the 1D shallow water equations, followed by adapting simplified climate models for quantum computers using only a few grid points, vertical levels, and prognostic variables.

Another, potentially near-term, use of quantum computers is to improve subgrid-scale parameterizations. Running QML-based parameterizations coupled to the climate model requires considerable quantum and classical runtime with overheads due to the quantum-classical coupling, since the quantum computer needs to run the QML parameterization. Recently proposed options that could help circumvent this challenge are surrogates or shadows of QML models (Schreiber et al., 2023; Jerbi et al., 2024). These are classical models emulating the outputs of previously trained QML models, thus potentially retaining some of the potential benefits while requiring a quantum device only during the training stage.

Tuning the free parameters of a climate model is a very time-consuming process that could be improved using quantum-assisted automatic routines. A first step towards developing an automatic calibration method could be tuning the Lorenz-96 model (Lorenz, 2006), which resembles the non-linear behavior of the climate system (Mouatadid et al., 2019).

Finally, for the analysis and evaluation of the resulting models, QML methods could provide potentially better classification of climate data, and alternative generative models efficiently reproduce realistic distributions of climate variables. The first steps for this would be to adapt the existing small-scale applications of quantum classifiers (Schuld and Killoran, 2019; Havlíček et al., 2019; Huang et al., 2021; Farhi and Neven, 2018; Abbas et al., 2021; Hur et al., 2022) and quantum generative models (Coyle et al., 2020; Dallaire-Demers and Killoran, 2018; Zoufal et al., 2019) to climate data, e.g., to simplified cloud classification problems.

All these research efforts are still in their infancy and face numerous challenges. Nevertheless, they offer exciting potential to significantly boost climate modelling to allow more robust regional climate projections and thus help manage the challenge posed by climate change, in addition to current efforts on hybrid ESMs (Eyring et al., 2024b) and high resolution modelling (Stevens et al., 2024).

**Open peer review.** To view the open peer review materials for this article, please visit http://doi.org/10.1017/eds.2025.10010.

**Author contribution.** Conceptualization: M.S., V.E., L.P. Visualization: L.P., M.S. Writing original draft: M.S., L.P., I.d.V., P.G., L.I., V.L., M.L., J.M.L., V.E. All authors approved the final submitted draft.

**Competing interests.** None.

**Data availability statement.** No data were used in this position paper.

**Ethics statement.** The research meets all ethical guidelines, including adherence to the legal requirements of the study country.

**Funding statement.** This project was made possible by the DLR Quantum Computing Initiative and the Federal Ministry for Economic Affairs and Climate Action; qci.dlr.de/projects/klim-qml. Funding for this study was also provided by the European Research Council (ERC) Synergy Grant "Understanding and Modelling the Earth System with Machine Learning (USMILE)" under the Horizon 2020 research and innovation programme (Grant agreement No. 855187). V.E. was additionally supported by the Deutsche Forschungsgemeinschaft (DFG, German Research Foundation) through the Gottfried Wilhelm Leibniz Prize awarded to Veronika Eyring (Reference No. EY 22/2-1). The research by L.I. and J.M.L. is partially funded in the context of the Munich Quantum Valley, which is supported by the Bavarian state government with funds from the Hightech Agenda Bayern Plus.

## References

**Aaronson S** (2015) Read the fine print. *Nature Physics 11*(4), 291–293.

**Abbas A**, **Sutter D**, **Zoufal C**, **Lucchi A**, **Figalli A and Woerner S** (2021) The power of quantum neural networks. *Nature Computational Science 1*, 403–409.

**Abel S**, **Blance A and Spannowsky M** (2022) Quantum optimization of complex systems with a quantum annealer. *Physical Review A 106*, 042607.

**Albash T and Lidar DA** (2018) Adiabatic quantum computation. *Reviews of Modern Physics 90*, 015002.

**Allen DR**, **Hoppel KW**, **Nedoluha GE**, **Eckermann SD and Barton CA** (2022) Ensemble-based gravity wave parameter retrieval for numerical weather prediction. *Journal of the Atmospheric Sciences 79*(3), 621–648.

**Alvanos M and Christoudias T** (2019) Accelerating atmospheric chemical kinetics for climate simulations. *IEEE Transactions on Parallel and Distributed Systems 30*(11), 2396–2407.

**Amin MH**, **Andriyash E**, **Rolfe J**, **Kulchytskyy B and Melko R** (2018) Quantum Boltzmann machine. *Physical Review X 8*, 021050.

**AQT** (2023) *Technical Report*. https://www.aqt.eu/ (accessed in April 2023).

**Arute F**, **Arya K**, **Babbush R**, **Bacon D**, **Bardin JC**, **Barends R**, **Biswas R**, **Boixo S**, **Brandao FGSL**, **Buell DA**, **Burkett B**, **Chen Y**, **Chen Z**, **Chiaro B**, **Collins R**, **Courtney W**, **Dunsworth A**, **Farhi E**, **Foxen B**, **Fowler A**, **Gidney C**, **Giustina M**, **Graff R**, **Guerin K**, **Habegger S**, **Harrigan MP**, **Hartmann MJ**, **Ho A**, **Hoffmann M**, **Huang T**, **Humble TS**, **Isakov SV**, **Jeffrey E**, **Jiang Z**, **Kafri D**, **Kechedzhi K**, **Kelly J**, **Klimov PV**, **Knysh S**, **Korotkov A**, **Kostritsa F**, **Landhuis D**, **Lindmark M**, **Lucero E**, **Lyakh D**, **Mandrà S**, **McClean JR**, **McEwen M**, **Megrant A**, **Mi X**, **Michielsen K**, **Mohseni M**, **Mutus J**, **Naaman O**, **Neeley M**, **Neill C**, **Niu MY**, **Ostby E**, **Petukhov A**, **Platt JC**, **Quintana C**, **Rieffel EG**, **Roushan P**, **Rubin NC**, **Sank D**, **Satzinger KJ**, **Smelyanskiy V**, **Sung KJ**, **Trevithick MD**, **Vainsencher A**, **Villalonga B**, **White T**, **Yao ZJ**, **Yeh P**, **Zalcman A**, **Neven H and Martinis JM** (2019) Quantum supremacy using a programmable superconducting processor. *Nature 574*, 505–510.

**Barz S**, **Kassal I**, **Ringbauer M**, **Lipp YO**, **Dakić B**, **Aspuru-Guzik A and Walther P** (2014) A two-qubit photonic quantum processor and its application to solving systems of linear equations. *Scientific Reports 4*, 6115.

**Bazgir A and Zhang Y** (2024) *QESM: A Leap Towards Quantum-Enhanced ML Emulation Framework for Earth and Climate Modeling*. https://doi.org/10.48550/arXiv.2410.01551.

**Behrens G**, **Beucler T**, **Gentine P**, **Iglesias-Suarez F**, **Pritchard M and Eyring V** (2022) Non-linear dimensionality reduction with a variational encoder decoder to understand convective processes in climate models. *Journal of Advances in Modeling Earth Systems 14*(8), e2022MS003130.

**Bellprat O**, **Kotlarski S**, **Lüthi D and Schär C** (2012) Objective calibration of regional climate models. *Journal of Geophysical Research, [Atmospheres] 117*(D23), D23115.

**Berry DW** (2014) High-order quantum algorithm for solving linear differential equations. *Journal of Physics A: Mathematical and Theoretical 47*(10), 105301.

**Besombes C**, **Pannekoucke O**, **Lapeyre C**, **Sanderson B and Thual O** (2021) Producing realistic climate data with generative adversarial networks. *Nonlinear Processes in Geophysics 28*(3), 347–370.

**Bharti K**, **Cervera-Lierta A**, **Kyaw TH**, **Haug T**, **Alperin-Lea S**, **Anand A**, **Degroote M**, **Heimonen H**, **Kottmann JS**, **Menke T**, **Mok W-K**, **Sim S**, **Kwek L-C and Aspuru-Guzik A** (2022) Noisy intermediate-scale quantum algorithms. *Reviews of Modern Physics 94*, 015004.

**Biamonte J**, **Wittek P**, **Pancotti N**, **Rebentrost P**, **Wiebe N and Lloyd S** (2017) Quantum machine learning. *Nature 549*, 195–202.

**Bock L**, **Lauer A**, **Schlund M**, **Barreiro M**, **Bellouin N**, **Jones C**, **Meehl GA**, **Predoi V**, **Roberts MJ and Eyring V** (2020) Quantifying progress across different CMIP phases with the ESMValTool. *Journal of Geophysical Research, [Atmospheres] 125*(21), e2019JD032321.

**Bondar DI**, **Gay-Balmaz F and Tronci C** (2019) Koopman wavefunctions and classical–quantum correlation dynamics. *Proceedings of the Royal Society A: Mathematical, Physical and Engineering Sciences 475*(2229), 20180879.

**Bonnet P**, **Pastori L**, **Schwabe M**, **Giorgetta MA**, **Iglesias-Suarez F and Eyring V** (2024) Tuning a climate model with machine-learning based emulators and history matching. *EGUsphere 2024*, 1–32.

**Bracco A**, **Brajard J**, **Dijkstra HA**, **Hassanzadeh P**, **Lessig C and Monteleoni C** (2025) Machine learning for the physics of climate. *Nature Reviews Physics 7*, 6–20 https://doi.org/10.1038/s42254-024-00776-3.

**Brassard G**, **Høyer P**, **Mosca M and Tapp A** (2002) Quantum Amplitude Amplification and Estimation. In *Quantum Computation and Information*, Lomonaco, Jr. SJ and Brandt, HE (eds) American Mathematical Society, Providence, Rhode Island, USA, pp. 53–74.

**Brenowitz ND**, **Beucler T**, **Pritchard M and Bretherton CS** (2020) Interpreting and stabilizing machine-learning parametrizations of convection. *Journal of the Atmospheric Sciences 77*(12), 4357–4375.

**Budinski L** (2021) Quantum algorithm for the advection–diffusion equation simulated with the lattice Boltzmann method. *Quantum Information Processing 20*(2), 57.

**Budinski L** (2022) Quantum algorithm for the Navier–Stokes equations by using the streamfunction-vorticity formulation and the lattice Boltzmann method. *International Journal of Quantum Information 20*(02), 2150039.

**Budinski L**, **Niemimäki O**, **Zamora-Zamora R and Lahtinen V** (2023) Efficient parallelization of quantum basis state shift. *Quantum Science and Technology 8*(4), 045031.

**Byrd GT and Ding Y** (2023) Quantum computing: Progress and innovation. *Computer 56*(1), 20–29.

**Cai X-D**, **Weedbrook C**, **Su Z-E**, **Chen M-C**, **Gu M**, **Zhu M-J**, **Li L**, **Liu N-L**, **Lu C-Y and Pan J-W** (2013) Experimental quantum computing to solve systems of linear equations. *Physical Review Letters 110*, 230501.

**Cai Z**, **Babbush R**, **Benjamin SC**, **Endo S**, **Huggins WJ**, **Li Y**, **McClean JR and O'Brien TE** (2023) Quantum Error Mitigation. *Reviews of Modern Physics 95*, 045005.

**Campbell ET**, **Terhal BM and Vuillot C** (2017) Roads towards fault-tolerant universal quantum computation. *Nature 549*(7671), 172–179.

**Caro MC**, **Huang H-Y**, **Cerezo M**, **Sharma K**, **Sornborger A**, **Cincio L and Coles PJ** (2022) Generalization in quantum machine learning from few training data. *Nature Communications 13*(1), 4919.

**Caro MC**, **Huang H-Y**, **Ezzell N**, **Gibbs J**, **Sornborger A**, **Cincio L**, **Coles PJ and Holmes Z** (2023) Out-of-distribution generalization for learning quantum dynamics. *Nature Communications 14*(1), 3751.

**Cerezo M**, **Arrasmith A**, **Babbush R**, **Benjamin SC**, **Endo S**, **Fujii K**, **McClean JR**, **Mitarai K**, **Yuan X**, **Cincio L and Coles PJ** (2021) Variational quantum algorithms. *Nature Reviews Physics 3*, 625–644.

**Cerezo M**, **Verdon G**, **Huang H-Y**, **Cincio L and Coles PJ** (2022) Challenges and opportunities in quantum machine learning. *Nature Computational Science 2*, 567–576.

**Cerezo M**, **Larocca M**, **García-Martín D**, **Diaz NL**, **Braccia P**, **Fontana E**, **Rudolph MS**, **Bermejo P**, **Ijaz A**, **Thanasilp S**, **Anschuetz ER and Holmes Z** (2023) Does provable absence of barren plateaus imply classical simulability? Or, why we need to rethink variational quantum computing. arXiv:2312.09121.

**Cichocki A**, **Lee N**, **Oseledets I**, **Phan A-H**, **Zhao Q and Mandic DP** (2016) Tensor networks for dimensionality reduction and large-scale optimization: Part 1 low-rank tensor decompositions. *Foundations and Trends in Machine Learning 9*(4-5), 249–429.

**Cinquegrana D**, **Zollo AL**, **Montesarchio M and Bucchignani E** (2023) A metamodel-based optimization of physical parameters of high resolution NWP ICON-LAM over southern Italy. *Atmosphere 14*(5), 788.

**Cong I**, **Choi S and Lukin MD** (2019) Quantum convolutional neural networks. *Nature Physics 15*, 1273–1278.

**Couvreux F**, **Hourdin F**, **Williamson D**, **Roehrig R**, **Volodina V**, **Villefranque N**, **Rio C**, **Audouin O**, **Salter J**, **Bazile E**, **Brient F**, **Favot F**, **Honnert R**, **Lefebvre M-P**, **Madeleine J-B**, **Rodier Q and Xu W** (2021) Process-based climate model development harnessing machine learning: I. A calibration tool for parameterization improvement. *Journal of Advances in Modeling Earth Systems 13*(3), e2020MS002217.

**Coyle B**, **Mills D**, **Danos V and Kashefi E** (2020) The born supremacy: Quantum advantage and training of an Ising born machine. *npj Quantum Information 6*, 60.

**D:Wave** (2023) *Technical Report*. https://www.dwavesys.com/ (accessed April 2023).

**Dallaire-Demers P-L and Killoran N** (2018) Quantum generative adversarial networks. *Physical Review A 98*, 012324.

**Das A and Chakrabarti BK** (2008) Colloquium: Quantum annealing and analog quantum computation. *Reviews of Modern Physics 80*, 1061–1081.

**Davenport JH**, **Jones JR and Thomason M** (2023) A practical overview of quantum computing: Is exascale possible? https://doi.org/10.48550/arXiv.2306.12346.

**Devitt SJ**, **Munro WJ and Nemoto K** (2013) Quantum error correction for beginners. *Reports on Progress in Physics 76*(7), 076001.

**Dilip R**, **Liu Y-J**, **Smith A and Pollmann F** (2022) Data compression for quantum machine learning. *Physical Review Research 4*, 043007.

**Du Y**, **Hsieh M-H**, **Liu T and Tao D** (2020) Expressive power of parametrized quantum circuits. *Physical Review Research 2*, 033125.

**Du Y**, **Hsieh M-H**, **Liu T**, **You S and Tao D** (2021) Learnability of quantum neural networks. *PRX Quantum 2*, 040337.

**Dunjko V and Briegel HJ** (2018) Machine learning & artificial intelligence in the quantum domain: A review of recent progress. *Reports on Progress in Physics 81*(7), 074001.

**Ebadi S**, **Keesling A**, **Cain M**, **Wang TT**, **Levine H**, **Bluvstein D**, **Semeghini G**, **Omran A**, **Liu J-G**, **Samajdar R**, **Luo X-Z**, **Nash B**, **Gao X**, **Barak B**, **Farhi E**, **Sachdev S**, **Gemelke N**, **Zhou L**, **Choi S**, **Pichler H**, **Wang S-T**, **Greiner M**, **Vuletić V and Lukin MD** (2022). Quantum optimization of maximum independent set using Rydberg atom arrays. *Science*, *376*(6598): 1209–1215.

**Eisert J**, **Cramer M and Plenio MB** (2010) Colloquium: Area laws for the entanglement entropy. *Reviews of Modern Physics 82*, 277–306.

**Enomoto Y**, **Anai K**, **Udagawa K and Takeda S** (2023) Continuous-variable quantum approximate optimization on a programmable photonic quantum processor. *Physical Review Research 5*, 043005.

**Eyring V**, **Gillett N**, **Rao KA**, **Barimalala R**, **Parrillo MB**, **Bellouin N**, **Cassou C**, **Durack P**, **Kosaka Y**, **McGregor S**, **Min S**, **Morgenstern O and Sun Y** (2021a) Human Influence on the Climate System. In *Climate change 2021: The physical science basis. Contribution of Working Group I to the sixth assessment report of the intergovernmental panel on climate change*. Masson-Delmotte V, Zhai P, Pirani A, Connors SL, Péan C, Berger S, Caud N, Chen Y, Goldfarb L, Gomis MI, Huang M, Leitzell K, Lonnoy E, Matthews JBR, Maycock TK, Waterfield T, Yelekçi O, Yu R, and Zhou B (eds). Cambridge, UK/New York, NY: Cambridge University Press, pp. 423–552.

**Eyring V**, **Mishra V**, **Griffith GP**, **Chen L**, **Keenan T**, **Turetsky MR**, **Brown S**, **Jotzo F**, **Moore FC and van der Linden S** (2021b) Reflections and projections on a decade of climate science. *Nature Climate Change 11*(4), 279–285.

**Eyring V**, **Collins WD**, **Gentine P**, **Barnes EA**, **Barreiro M**, **Beucler T**, **Bocquet M**, **Bretherton CS**, **Christensen HM**, **Dagon K**, **Gagne DJ**, **Hall D**, **Hammerling D**, **Hoyer S**, **Iglesias-Suarez F**, **Lopez-Gomez I**, **McGraw MC**, **Meehl GA**, **Molina MJ**, **Monteleoni C**, **Mueller J**, **Pritchard MS**, **Rolnick D**, **Runge J**, **Stier P**, **Watt-Meyer O**, **Weigel K**, **Yu R and Zanna L** (2024a) Pushing the frontiers in climate modelling and analysis with machine learning. *Nature Climate Change 14*(9), 916–928.

**Eyring V**, **Gentine P**, **Camps-Valls G**, **Lawrence DM and Reichstein M** (2024b) AI-empowered next-generation multiscale climate modeling for mitigation and adaptation. *Nature Geoscience 17*(10), 963–971.

**Farhi E and Neven H** (2018) Classification with quantum neural networks on near term processors. arXiv:1802.06002.

**Farhi E**, **Goldstone J and Gutmann S** (2014) A quantum approximate optimization algorithm. arXiv:1411.4028

**Ferreira da Silva R**, **Badia RM**, **Bard D**, **Foster IT**, **Jha S and Suter F** (2024) Frontiers in scientific workflows: Pervasive integration with high-performance computing. *Computer 57*(8), 36–44.

**Gaitan F** (2020) Finding flows of a Navier–Stokes fluid through quantum computing. *npj Quantum Information 6*(1), 61.

**Gentine P**, **Eyring V and Beucler T** (2021) Deep learning for the parametrization of subgrid processes in climate models. In Camps-Valls G, Tuia D, Zhu XX and Reichstein M (eds.), *Deep Learning for the Earth Sciences* (2nd edn). Wiley, Hoboken, NJ, USA, pp. 307–314.

**Gettelman A and Rood RB** (2016) *Demystifying Climate Models*. Berlin/Heidelberg: Springer.

**Google Quantum AI** (2023) *Technical Report*. https://quantumai.google/hardware (accessed April 2023).

**Google Quantum AI and Collaborators** (2025) Quantum error correction below the surface code threshold. *Nature 638*, 920–926.

**Grumbling E and Horowitz M** (eds) (2019) In *Quantum Computing: Progress and Prospects*. Washington, DC: The National Academies Press.

**Grundner A**, **Beucler T**, **Gentine P and Eyring V** (2024) Data-driven equation discovery of a cloud cover parameterization. *Journal of Advances in Modeling Earth Systems 16*(3), e2023MS003763.

**Hangleiter D and Eisert J** (2023) Computational advantage of quantum random sampling. *Reviews of Modern Physics 95*, 035001.

**Harrow AW**, **Hassidim A and Lloyd S** (2009) Quantum algorithm for linear systems of equations. *Physical Review Letters 103*, 150502.

**Havlíček V**, **Córcoles AD**, **Temme K**, **Harrow AW**, **Kandala A**, **Chow JM and Gambetta JM** (2019) Supervised learning with quantum-enhanced feature spaces. *Nature 567*, 209–212.

**Hohenegger C**, **Kornblueh L**, **Klocke D**, **Becker T**, **Cioni G**, **Engels JF**, **Schulzweida U and Stevens B** (2020) Climate statistics in global simulations of the atmosphere, from 80 to 2.5 km grid spacing. *Journal of the Meteorological Society of Japan Ser II 98*(1), 73–91.

**Hourdin F**, **Mauritsen T**, **Gettelman A**, **Golaz J-C**, **Balaji V**, **Duan Q**, **Folini D**, **Ji D**, **Klocke D**, **Qian Y**, **Rauser F**, **Rio C**, **Tomassini L**, **Watanabe M and Williamson D** (2017) The art and science of climate model tuning. *Bulletin of the American Meteorological Society 98*(3), 589–602.

**Huang H-Y**, **Kueng R and Preskill J** (2020) Predicting many properties of a quantum system from very few measurements. *Nature Physics 16*, 1050–1057.

**Huang H-Y**, **Broughton M**, **Mohseni M**, **Babbush R**, **Boixo S**, **Neven H and McClean JR** (2021) Power of data in quantum machine learning. *Nature Communications 12*(1), 2631.

**Humble TS**, **McCaskey A**, **Lyakh DI**, **Gowrishankar M**, **Frisch A and Monz T** (2021) Quantum computers for high-performance computing. *IEEE Micro 41*(5), 15–23.

**Hur T**, **Kim L and Park DK** (2022) Quantum convolutional neural network for classical data classification. *Quantum Machine Intelligence 4*, 3.

**IBM** (2023) *Technical Report*. https://www.ibm.com/roadmaps/quantum.pdf (accessed April 2023).

**IonQ** (2023) *Technical Report*. https://ionq.com/ (accessed April 2023).

**Itani W and Succi S** (2022) Analysis of Carleman linearization of lattice Boltzmann. *Fluids 7*(1), 24.

**Izawa S**, **Kitai K**, **Tanaka S**, **Tamura R and Tsuda K** (2022) Continuous black-box optimization with an Ising machine and random subspace coding. *Physical Review Research 4*, 023062.

**Jacobson MZ** (2005) *Fundamentals of Atmospheric Modeling*. Cambridge University Press, Cambridge, UK.

**Jaderberg B**, **Gentile AA**, **Ghosh A**, **Elfving VE**, **Jones C**, **Vodola D**, **Manobianco J and Weiss H** (2024) Potential of quantum scientific machine learning applied to weather modelling. arXiv:2404.08737

**Jerbi S**, **Gyurik C**, **Marshall SC**, **Molteni R and Dunjko V** (2024) Shadows of quantum machine learning. *Nature Communications 15*, 5676.

**Joshi MK**, **Kranzl F**, **Schuckert A**, **Lovas I**, **Maier C**, **Blatt R**, **Knap M and Roos CF** (2022) Observing emergent hydrodynamics in a long-range quantum magnet. *Science 376*(6594), 720–724.

**Kacewicz B** (2006) Almost optimal solution of initial-value problems by randomized and quantum algorithms. *Journal of Complexity 22*(5), 676–690.

**Kaps A**, **Lauer A**, **Camps-Valls G**, **Gentine P**, **Gomez-Chova L and Eyring V** (2023) Machine-learned cloud classes from satellite data for process-oriented climate model evaluation. *IEEE Transactions on Geoscience and Remote Sensing 61*, 1–15.

**King AD**, **Nocera A**, **Rams MM**, **Dziarmaga J**, **Wiersema R**, **Bernoudy W**, **Raymond J**, **Kaushal N**, **Heinsdorf N**, **Harris R**, **Boothby K**, **Altomare F**, **Berkley AJ**, **Boschnak M**, **Chern K**, **Christiani H**, **Cibere S**, **Connor J**, **Dehn MH**, **Deshpande R**, **Ejtemaee S**, **Farré P**, **Hamer K**, **Hoskinson E**, **Huang S**, **Johnson MW**, **Kortas S**, **Ladizinsky E**, **Lai T**, **Lanting T**, **Li R**, **MacDonald AJR**, **Marsden G**, **McGeoch CC**, **Molavi R**, **Neufeld R**, **Norouzpour M**, **Oh T**, **Pasvolsky J**, **Poitras P**, **Poulin-Lamarre G**, **Prescott T**, **Reis M**, **Rich C**, **Samani M**, **Sheldan B**, **Smirnov A**, **Sterpka E**, **Clavera BT**, **Tsai N**, **Volkmann M**, **Whiticar A**, **Whittaker JD**, **Wilkinson W**, **Yao J**, **Yi TJ**, **Sandvik AW**, **Alvarez G**, **Melko RG**, **Carrasquilla J**, **Franz M and Amin MH** (2024) Computational supremacy in quantum simulation. arXiv:2403.00910

**Koh YW and Nishimori H** (2022) Quantum and classical annealing in a continuous space with multiple local minima. *Physical Review A 105*, 062435.

**Lau JWZ**, **Lim KH**, **Shrotriya H and Kwek LC** (2022) NISQ computing: Where are we and where do we go? *AAPPS Bulletin 32*(1), 27.

**Li X**, **Yin X**, **Wiebe N**, **Chun J**, **Schenter GK**, **Cheung MS and Mülmenstädt J** (2023) Potential quantum advantage for simulation of fluid dynamics. arXiv:2303.16550

**Lloyd S**, **Palma GD**, **Gokler C**, **Kiani B**, **Liu Z-W**, **Marvian M**, **Tennie F and Palmer T** (2020) Quantum algorithm for nonlinear differential equations. arXiv:2011.06571.

**Lorenz EN** (2006) Predictability—A Problem Partly Solved. In: Palmer T, Hagedorn R (eds) *Predictability of Weather and Climate*. Cambridge: Cambridge University Press, pp. 40–58.

**Madsen LS**, **Laudenbach F**, **Askarani MF**, **Rortais F**, **Vincent T**, **Bulmer JFF**, **Miatto FM**, **Neuhaus L**, **Helt LG**, **Collins MJ**, **Lita AE**, **Gerrits T**, **Nam SW**, **Vaidya VD**, **Menotti M**, **Dhand I**, **Vernon Z**, **Quesada N and Lavoie J** (2022) Quantum computational advantage with a programmable photonic processor. *Nature 606*(7912), 75–81.

**Markidis S** (2022) On physics-informed neural networks for quantum computers. *Frontiers in Applied Mathematics and Statistics 8*, 1036711.

**Maronese M**, **Moro L**, **Rocutto L and Prati E** (2022) *Quantum Compiling*. Cham: Springer International Publishing, pp. 39–74.

**Masson-Delmotte V**, **Zhai P**, **Pirani A**, **Connors S**, **Péan C**, **Berger S**, **Caud N**, **Chen Y**, **Goldfarb L**, **Gomis MI**, **Huang M**, **Leitzell K**, **Lonnoy E and Matthews JBR** (2021) Climate change 2021: The physical science basis. In Maycock TK, Waterfield T, Yelekçi O and Zhou B (eds.), *Contribution of Working Group I to the Sixth Assessment Report of the Intergovernmental Panel on Climate Change (IPCC)*. Cambridge University Press, Cambridge, UK/ New York, NY.

**Meyer JJ**, **Mularski M**, **Gil-Fuster E**, **Mele AA**, **Arzani F**, **Wilms A and Eisert J** (2023) Exploiting symmetry in variational quantum machine learning. *PRX Quantum 4*(1), 010328.

**Mezzacapo A**, **Sanz M**, **Lamata L**, **Egusquiza IL**, **Succi S and Solano E** (2015) Quantum simulator for transport phenomena in fluid flows. *Scientific Reports 5*(1), 13153.

**Mouatadid S**, **Gentine P**, **Yu W and Easterbrook S** (2019) Recovering the parameters underlying the Lorenz-96 chaotic dynamics. arXiv:1906.06786.

**Mücke S**, **Heese R**, **Müller S**, **Wolter M and Piatkowski N** (2023) Feature selection on quantum computers. *Quantum Machine Intelligence 5*, 11.

**Neumann P**, **Düben P**, **Adamidis P**, **Bauer P**, **Brück M**, **Kornblueh L**, **Klocke D**, **Stevens B**, **Wedi N and Biercamp J** (2019) Assessing the scales in numerical weather and climate predictions: Will exascale be the rescue? *Philosophical Transactions of the Royal Society of London, Series A: Mathematical, Physical and Engineering Sciences 377*(2142), 20180148.

**Nielsen MA and Chuang IL** (2010) *Quantum Computation and Quantum Information: 10th Anniversary Edition*. Cambridge University Press, Cambridge, UK.

**Nivelkar M**, **Bhirud S**, **Singh M**, **Ranjan R and Kumar B** (2023) Quantum computing to study cloud turbulence properties. *IEEE Access 11*, 70679–70690.

**Otgonbaatar S**, **Nurmi O**, **Johansson M**, **Mäkelä J**, **Kocman T**, **Gawron P**, **Puchała Z**, **Mielzcarek J**, **Miroszewski A and Dumitru CO** (2023). Quantum computing for climate change detection, climate modeling, and climate digital twin. https://doi.org/10.36227/techrxiv.24478663.v1.

**Otten M**, **Goumiri IR**, **Priest BW**, **Chapline GF and Schneider MD** (2020) Quantum machine learning using Gaussian processes with performant quantum kernels. arXiv:2004.11280.

**Pahlavan HA**, **Hassanzadeh P and Alexander MJ** (2024) Explainable offline-online training of neural networks for parameterizations: A 1d gravity wave-QBO testbed in the small-data regime. *Geophysical Research Letters 51*(2), e2023GL106324.

**PASQAL** (2023) *Technical Report*. https://www.pasqal.com/ (accessed April 2023).

**Pastori L**, **Grundner A**, **Eyring V and Schwabe M** (2025) Quantum neural network based cloud cover parameterization for climate models. arXiv:2502.10131. https://arxiv.org/abs/2502.10131.

**Pérez-Salinas A**, **Cervera-Lierta A**, **Gil-Fuster E and Latorre JI** (2020) Data re-uploading for a universal quantum classifier. *Quantum 4*, 226.

**Peruzzo A**, **McClean J**, **Shadbolt P**, **Yung M-H**, **Zhou X-Q**, **Love PJ**, **Aspuru-Guzik A and O'Brien JL** (2014) A variational eigenvalue solver on a photonic quantum processor. *Nature Communications 5*, 4213.

**Pesah A**, **Cerezo M**, **Wang S**, **Volkoff T**, **Sornborger AT and Coles PJ** (2021) Absence of barren plateaus in quantum convolutional neural networks. *Physical Review X 11*, 041011.

**Preskill J** (2018) Quantum computing in the NISQ era and beyond. *Quantum 2*, 79.

**Pudenz KL**, **Albash T and Lidar DA** (2014) Error-corrected quantum annealing with hundreds of qubits. *Nature Communications 5*(1), 3243.

**Quantinuum** (2023) *Technical Report*. https://www.quantinuum.com/ (accessed April 2023).

**Ragone M**, **Bakalov BN**, **Sauvage F**, **Kemper AF**, **Marrero CO**, **Larocca M and Cerezo M** (2024) A Lie algebraic theory of barren plateaus for deep parameterized quantum circuits. *Nature Communications 15*, 7172. https://doi.org/10.1038/s41467-024-49909-3

**Rahman SM**, **Alkhalaf OH**, **Alam MS**, **Tiwari SP**, **Shafiullah M**, **Al-Judaibi SM and Al-Ismail FS** (2024) Climate change through quantum lens: Computing and machine learning. *Earth Systems and Environment 8*(3), 705–722.

**Rapp F and Roth M** (2024) Quantum Gaussian process regression for Bayesian optimization. *Quantum Machine Intelligence 6*, 5.

**Rasp S**, **Pritchard MS and Gentine P** (2018) Deep learning to represent subgrid processes in climate models. *Proceedings of the National Academy of Sciences of the United States of America 115*(39), 9684–9689.

**Ray N**, **Banerjee T**, **Nadiga B and Karra S** (2019) Towards solving the Navier-Stokes equation on quantum computers. arXiv:1904.09033.

**rigetti** (2023) *Technical Report*. https://www.rigetti.com/ (accessed April 2023).

**Rudolph MS**, **Lerch S**, **Thanasilp S**, **Kiss O**, **Vallecorsa S**, **Grossi M and Holmes Z** (2023) Trainability barriers and opportunities in quantum generative modeling. *npj Quantum Information 10*(1), 116. arXiv:2305.02881.

**Rüfenacht M**, **Taketani B**, **Lähteenmäki P**, **Bergholm V**, **Kranzlmüller D**, **Schulz L and Schulz M** (2022) *Bringing Quantum Acceleration to Supercomputers. White Paper*. https://meetiqm.com/uploads/documents/IQM_HPC-QC-Integration-Whitepaper.pdf.

**Sander R**, **Kerkweg A**, **Jöckel P and Lelieveld J** (2005) Technical note: The new comprehensive atmospheric chemistry module mecca. *Atmospheric Chemistry and Physics 5*, 445–450.

**Schalkers MA and Möller M** (2022) Efficient and fail-safe collisionless quantum Boltzmann method. *Journal of Computational Physics 502*, 11286. arXiv:2211.14269.

**Schär C**, **Fuhrer O**, **Arteaga A**, **Ban N**, **Charpilloz C**, **Girolamo SD**, **Hentgen L**, **Hoefler T**, **Lapillonne X**, **Leutwyler D**, **Osterried K**, **Panosetti D**, **Rüdisühli S**, **Schlemmer L**, **Schulthess TC**, **Sprenger M**, **Ubbiali S and Wernli H** (2020) Kilometer-scale climate models: Prospects and challenges. *Bulletin of the American Meteorological Society 101*(5), E567–E587.

**Scholl P**, **Schuler M**, **Williams HJ**, **Eberharter AA**, **Barredo D**, **Schymik K-N**, **Lienhard V**, **Henry L-P**, **Lang TC**, **Lahaye T**, **Läuchli AM and Browaeys A** (2021) Quantum simulation of 2d antiferromagnets with hundreds of rydberg atoms. *Nature 595*(7866), 233–238.

**Schreiber FJ**, **Eisert J and Meyer JJ** (2023) Classical surrogates for quantum learning models. *Physical Review Letters 131*, 100803.

**Schuld M and Killoran N** (2019) Quantum machine learning in feature Hilbert spaces. *Physical Review Letters 122*, 040504.

**Schuld M and Petruccione F** (2018) *Supervised Learning with Quantum Computers*. Cham: Springer.

**Schuld M**, **Sinayskiy I and Petruccione F** (2015) An introduction to quantum machine learning. *Contemporary Physics 56*(2), 172–185.

**Schuld M**, **Sweke R and Meyer JJ** (2021) Effect of data encoding on the expressive power of variational quantum-machine-learning models. *Physical Review A 103*, 032430.

**Schulz M**, **Kranzlmüller D**, **Schulz LB**, **Trinitis C and Weidendorfer J** (2021) On the inevitability of integrated HPC systems and how they will change HPC system operations. In *HEART '21: Proceedings of the 11th International Symposium on Highly Efficient Accelerators and Reconfigurable Technologies*. New York. Association for Computing Machinery.

**Schulz M**, **Ruefenacht M**, **Kranzlmuller D and Schulz L** (2022) Accelerating HPC with quantum computing: It is a software challenge too. *Computing in Science & Engineering 24*(04), 60–64.

**Semeghini G, Levine H, Keesling A, Ebadi S, Wang TT, Bluvstein D, Verresen R, Pichler H, Kalinowski M, Samajdar R, Omran A, Sachdev S, Vishwanath A, Greiner M, Vuletić V and Lukin MD** (2021). Probing topological spin liquids on a programmable quantum simulator. *Science 374*(6572), 1242–1247.

**Sevilla J and Riedel CJ** (2020) *Forecasting Timelines of Quantum Computing*. arXiv:2009.05045.

**Sherwood SC, Bony S and Dufresne J-L** (2014) Spread in model climate sensitivity traced to atmospheric convective mixing. *Nature 505*(7481), 37–42.

**Shokri N, Stevens B, Madani K, Grabe J, Schlüter M and Smirnova I** (2022) Climate informed engineering: An essential pillar of industry 4.0 transformation. *ACS Engineering Au Journal 3*(1), 3–6.

**Singh M, Dhara C, Kumar A, Gill S and Uhlig S** (2022) *Artificial intelligence, machine learning and Blockchain in Quantum satellite, drone and network*. In *Chapter Quantum Artificial Intelligence for the Science of Climate Change*. CRC Press, pp. 199–207.

**Steijl R and Barakos GN** (2018) Parallel evaluation of quantum algorithms for computational fluid dynamics. *Computers and Fluids 173*, 22–28.

**Stevens B, Satoh M, Auger L, Biercamp J, Bretherton CS, Chen X, Düben P, Judt F, Khairoutdinov M, Klocke D, Kodama C, Kornblueh L, Lin S-J, Neumann P, Putman WM, Röber N, Shibuya R, Vanniere B, Vidale PL, Wedi N and Zhou L** (2019) DYAMOND: The dynamics of the atmospheric general circulation modeled on non-hydrostatic domains. *Progress in Earth and Planetary Science 6*(1), 61.

**Stevens B, Adami S, Ali T, Anzt H, Aslan Z, Attinger S, Bäck J, Baehr J, Bauer P, Bernier N, Bishop B, Bockelmann H, Bony S, Brasseur G, Bresch DN, Breyer S, Brunet G, Buttigieg PL, Cao J, Castet C, Cheng Y, Dey Choudhury A, Coen D, Crewell S, Dabholkar A, Dai Q, Doblas-Reyes F, Durran D, El Gaidi A, Ewen C, Exarchou E, Eyring V, Falkinhoff F, Farrell D, Forster PM, Frassoni A, Frauen C, Fuhrer O, Gani S, Gerber E, Goldfarb D, Grieger J, Gruber N, Hazeleger W, Herken R, Hewitt C, Hoefler T, Hsu H-H, Jacob D, Jahn A, Jakob C, Jung T, Kadow C, Kang I-S, Kang S, Kashinath K, Kleinen-von Königslöw K, Klocke D, Kloenne U, Klöwer M, Kodama C, Kollet S, Kölling T, Kontkanen J, Kopp S, Koran M, Kulmala M, Lappalainen H, Latifi F, Lawrence B, Lee JY, Lejeun Q, Lessig C, Li C, Lippert T, Luterbacher J, Manninen P, Marotzke J, Matsouoka S, Merchant C, Messmer P, Michel G, Michielsen K, Miyakawa T, Müller J, Munir R, Narayanasetti S, Ndiaye O, Nobre C, Oberg A, Oki R, Özkan Haller T, Palmer T, Posey S, Prein A, Primus O, Pritchard M, Pullen J, Putrasahan D, Quaas J, Raghavan K, Ramaswamy V, Rapp M, Rauser F, Reichstein M, Revi A, Saluja S, Satoh M, Schemann V, Schemm S, Schnadt Poberaj C, Schulthess T, Senior C, Shukla J, Singh M, Slingo J, Sobel A, Solman S, Spitzer J, Stier P, Stocker T, Strock S, Su H, Taalas P, Taylor J, Tegtmeier S, Teutsch G, Tompkins A, Ulbrich U, Vidale P-L, Wu C-M, Xu H, Zaki N, Zanna L, Zhou T and Ziemen F** (2024) Earth virtualization engines (eve). *Earth System Science Data 16*(4), 2113–2122.

**Struchalin G, Zagorovskii YA, Kovlakov E, Straupe S and Kulik S** (2021) Experimental estimation of quantum state properties from classical shadows. *PRX Quantum 2*(1), 010307.

**Tennie F and Palmer TN** (2023) Quantum computers for weather and climate prediction: The good, the bad and the noisy. *Bulletin of the American Meteorological Society 104*(2), E488–E500.

**Thanasilp S, Wang S, Cerezo M and Holmes Z** (2022) Exponential concentration and untrainability in quantum kernel methods. *Nature Communications 15*. arXiv:2208.11060.

**Watson-Parris D, Williams A, Deaconu L and Stier P** (2021) Model calibration using ESEm v1.1.0 – An open, scalable Earth system emulator. *Geoscientific Model Development 14*(12), 7659–7672.

**Wu Y, Bao W-S, Cao S, Chen F, Chen M-C, Chen X, Chung T-H, Deng H, Du Y, Fan D, Gong M, Guo C, Guo C, Guo S, Han L, Hong L, Huang H-L, Huo Y-H, Li L, Li N, Li S, Li Y, Liang F, Lin C, Lin J, Qian H, Qiao D, Rong H, Su H, Sun L, Wang L, Wang S, Wu D, Xu Y, Yan K, Yang W, Yang Y, Ye Y, Yin J, Ying C, Yu J, Zha C, Zhang C, Zhang H, Zhang K, Zhang Y, Zhao H, Zhao Y, Zhou L, Zhu Q, Lu C-Y, Peng C-Z, Zhu X and Pan J-W** (2021) Strong quantum computational advantage using a superconducting quantum processor. *Physical Review Letters 127*, 180501.

**Yarkoni S, Raponi E, Bäck T and Schmitt S** (2022) Quantum annealing for industry applications: Introduction and review. *Reports on Progress in Physics 85*(10), 104001.

**Yu S, Hannah W, Peng L, Lin J, Bhouri MA, Gupta R, Lütjens B, Will JC, Behrens G, Busecke J, Loose N, Stern C, Beucler T, Harrop B, Hillman B, Jenney A, Ferretti SL, Liu N, Anandkumar A, Brenowitz N, Eyring V, Geneva N, Gentine P, Mandt S, Pathak J, Subramaniam A, Vondrick C, Yu R, Zanna L, Zheng T, Abernathey R, Ahmed F, Bader D, Baldi P, Barnes E, Bretherton C, Caldwell P, Chuang W, Han Y, Huang Y, Iglesias-Suarez F, Jantre S, Kashinath K, Khairoutdinov M, Kurth T, Lutsko N, Ma P-L, Mooers G, Neelin JD, Randall D, Shamekh S, Taylor M, Urban N, Yuval J, Zhang G and Pritchard M** (2023a) Climsim: A large multi-scale dataset for hybrid physics-ml climate emulation. In Oh A, Naumann T, Globerson A, Saenko K, Hardt M and Levine S (eds.), *Advances in Neural Information Processing Systems* (Vol. *36*). Curran Associates, Inc, Red Hook, NY, USA, pp. 22070–22084.

**Yu Z, Chen Q, Jiao Y, Li Y, Lu X, Wang X and Yang JZ** (2023b) *Provable Advantage of Parameterized Quantum Circuit in Function Approximation*. *arXiv:2310.07528*.

**Zhang T, Li L, Lin Y, Xue W, Xie F, Xu H and Huang X** (2015) An automatic and effective parameter optimization method for model tuning. *Geoscientific Model Development 8*(11), 3579–3591.

**Zhang K, Liu L, Hsieh M-H and Tao D** (2022) Escaping from the barren plateau via Gaussian initializations in deep variational quantum circuits. In Koyejo S, Mohamed S, Agarwal A, Belgrave D, Cho K, Oh A (eds) *Proceedings of the 36th International*

*Conference on Neural Information Processing Systems (NIPS '22)*. Curran Associates Inc., Red Hook, NY, USA, Article 1352, 18612–18627.

**Zheng Y**, **Song C**, **Chen M-C**, **Xia B**, **Liu W**, **Guo Q**, **Zhang L**, **Xu D**, **Deng H**, **Huang K**, **Wu Y**, **Yan Z**, **Zheng D**, **Lu L**, **Pan J-W**, **Wang H**, **Lu C-Y and Zhu X** (2017) Solving systems of linear equations with a superconducting quantum processor. *Physical Review Letters 118*, 210504.

**Zhong H-S**, **Wang H**, **Deng Y-H**, **Chen M-C**, **Peng L-C**, **Luo Y-H**, **Qin J**, **Wu D**, **Ding X**, **Hu Y**, **Hu P**, **Yang X-Y**, **Zhang W-J**, **Li H**, **Li Y**, **Jiang X**, **Gan L**, **Yang G**, **You L**, **Wang Z**, **Li L**, **Liu N-L**, **Lu C-Y and Pan J-W** (2020) Quantum computational advantage using photons. *Science 370*(6523), 1460–1463.

**Zhu Q**, **Cao S**, **Chen F**, **Chen M-C**, **Chen X**, **Chung T-H**, **Deng H**, **Du Y**, **Fan D**, **Gong M**, **Guo C**, **Guo C**, **Guo S**, **Han L**, **Hong L**, **Huang H-L**, **Huo Y-H**, **Li L**, **Li N**, **Li S**, **Li Y**, **Liang F**, **Lin C**, **Lin J**, **Qian H**, **Qiao D**, **Rong H**, **Su H**, **Sun L**, **Wang L**, **Wang S**, **Wu D**, **Wu Y**, **Xu Y**, **Yan K**, **Yang W**, **Yang Y**, **Ye Y**, **Yin J**, **Ying C**, **Yu J**, **Zha C**, **Zhang C**, **Zhang H**, **Zhang K**, **Zhang Y**, **Zhao H**, **Zhao Y**, **Zhou L**, **Lu C-Y**, **Peng C-Z**, **Zhu X, and Pan J-W** (2022). Quantum computational advantage via 60-qubit 24-cycle random circuit sampling. *Science Bulletin 67*(3):240–245.

**Zlatev Z**, **Dimov I**, **Faragó I**, **Georgiev K and Havasi Á** (2022) Running an atmospheric chemistry scheme from a large air pollution model by using advanced versions of the Richardson extrapolation. In Lirkov, I., Margenov, S. (eds) *Large-Scale Scientific Computing LSSC 2021. Lecture Notes in Computer Science*, vol *13127*. Springer, Cham., pp 188–197.

**Zoufal C**, **Lucchi A and Woerner S** (2019) Quantum generative adversarial networks for learning and loading random distributions. *npj Quantum Information 5*, 103.